\documentclass[conference]{IEEEtran}
\ifCLASSINFOpdf
\else
\fi
\usepackage{caption}
\usepackage{subcaption}
\usepackage{graphicx}
\usepackage{color}
\usepackage{amsmath}
\usepackage{amssymb}
\usepackage{color}
\usepackage{epstopdf}
\usepackage{cite}
\usepackage{array}
\usepackage{multirow}
\usepackage{hhline}
\usepackage{graphicx}
\usepackage{epsfig}
\usepackage{epstopdf}
\usepackage{cite}
\usepackage{graphicx}
\usepackage{latexsym}
\usepackage{amsmath}
\usepackage{amssymb}
\usepackage{mathrsfs}
\usepackage{float}
\usepackage{epsfig}
\usepackage{cite}
\usepackage{amssymb}
\usepackage[english]{babel}
\usepackage{array}
\usepackage{color}
\usepackage[utf8]{inputenc}
\usepackage{pbox}
\usepackage{booktabs}
\usepackage{longtable}
\usepackage{bigstrut}
\usepackage{tabularx}
\usepackage{tabulary} 

\usepackage{hyperref} 
\begin{document}
\title{Review of isolation enhancement with the help of Theory of characteristic modes}

\author{Farhan Ammar Ahmad\\ Department of Electrical Engineering\\ University of of Management and Technology Lahore,\\ Sialkot Campus, Pakistan\\ farhan.sayal@post.umt.edu.pk}


\markboth{Journal of \LaTeX\ Class Files,~Vol.~14, No.~8, August~2015}%
{Shell \MakeLowercase{\textit{et al.}}: Bare Demo of IEEEtran.cls for IEEE Journals}
\maketitle
\begin{abstract}
Multiple-input-multiple-output (MIMO) antennas performance can be degraded due to the poor isolation between the MIMO antenna elements. In this paper a review of the different isolation enhancement schemes available in the literature is presented. Empirically the isolation between the antennas can be improved by placing the antenna as far as possible and it can be enhanced further by introducing different isolation enhancement schemes. Theory of characteristic modes (TCM) was recently proposed that had useful benefits. TCM was also used to enhance the isolation. This papers will also focus on the different approaches of TCM, to enhance the isolation. 

\end{abstract}


\IEEEpeerreviewmaketitle

\section{Introduction}
\IEEEPARstart{T}{heory} of characteristic modes (TCM) was developed by Garbacez \cite{Garbacz} but it gained importance after it was revisited in \cite{Harrington} by diagonalizing the impedance matrix of the body. Mathematically \cite{cabedo_thesis,eva_thesis,asim_thesis,Asim_Mecap_2}
\begin{equation}
[Z]I_n = (1+j\lambda_n)[R]I_n 
\label{2_TCM_equ2}
\end{equation}
Where $Z$ represents the impedance and $Z=R+jX$, $I_n$ represents the current eigen vector and $1+j\lambda_n$ represents the eigen value. Further simplification of the (\ref{2_TCM_equ2}) gives

\begin{equation}
[X]I_n = \lambda_n[R]I_n
\end{equation}
It can be also represented as 
\begin{equation}
[X]J_n = \lambda_n[R]J_n
\label{equ_2}
\end{equation}

Here $J_n$ represents the current density eigen vector. From Theorem of Reciprocity we know that, If $Z$ is a linear symmetric operator then the Hermitian parts of of $Z$ ($R$ and $X$) will be also linear symmetric operator \cite{cabedo_thesis}. Thus we can conclude that the eigen values will be always real and we can assume that the eigen vectors will always be real and equiphasal. Eigen value gives us an information about the behavior of a modes at a particular frequency whether it will resonate or store electrical energy or store mechanical energy. Modal Significance (MS) is a parameter depending on the eigen value and gives us an information about the contribution of a particular mode at a particular frequency. Mathematically it is given by
\begin{equation}
MS_n = \left |\dfrac{1}{1+j \lambda_n} \right |
\label{MS}
\end{equation}

At resonance the value of MS is equal to 1 and the 3 dB BW corresponds to 0.707 value. The radiation pattern associated with these real eigen modes are orthogonal to one another.

TCM is widely used for the  analysis  of  various  different  types  of  antennas  such  as  wire  antennas  \cite{21},  handset  antennas  \cite{22},  small  antennas  \cite{23},  dielectric  antennas  \cite{24},  printed  MIMO  antennas \cite{DGS9,Asim_Access,Asim_FERMAT,Asim_DGS_mecap_1,Asim_AWPL} and  slot  antennas 	\cite{Asim_MMCE,Asim_APS_2017_1,Asim_EuCAP_2017_2}

Multiple-input-multiple-output (MIMO) antennas are widely used for the enhancement of antenna capacity and this technology will be used for 4G as well as 5G communications \cite{sharawi_book,Asim_SIW}. The performance of such antennas can be degraded by if the antennas are not properly isolated \cite{Asim_RWG,sharawi_book}. Recently TCM was used to enhance the isolation between the MIMO antenna elements.

The main contribution of this paper will be to review the empirical method procedure (EMP) used for the isolation enhancement as well as we will review the isolation enhancement from the perspective of TCM. The short comings as well as the significance of the methods proposed in the literature w.r.t TCM will be highlighted. Section II will briefly discuss the different methods present in the literature from EMP method while section III will discuss isolation enhancement from the perspective of TCM. The conclusion is provided in section IV.

\section{Isolation Enhancement using EMP}
From EMP perspective the isolation between MIMO antennas can be enhanced by the use of meta surfaces or electromagnetic bandgap structures \cite{Meta1,Meta2,Meta3}, parasitic elements \cite{PE1,PE2}, neutralization line technique \cite{NL1,NL2},  optimization of the antenna system configuration \cite{AOP1,AOP2}, decoupling networks \cite{DN1}, and the use of Defected ground structures (DGS)\cite{DGS1,DGS2,DGS3,DGS4,DGS5,DGS6,DGS7,DGS8,DGS9,DGS10,DGS11}. Among all the aforementioned port isolation enhancement methods, DGS is the least complex and expensive.

Meta-material (Single layer or Double layer) are used for the miniaturization as well as for isolation. They suppress the surface waves between the antennas resulting in an enhanced isolation \cite{Meta1,Meta2,Meta3}. Meta-materials normally involve larger and bulky formations that needs to be properly optimized. The isolation enhancement cant be obtained easily, if miniaturization is the major objective. Isolation improvement in \cite{DN1} by the use of decoupling methods normally compromises of two stages; matching network and then a decoupling network. In decoupling network reactive elements are designed between the antenna feeding ports to improve the isolation, such networks require more space and generates additional cost. In addition neutralization line technique was proposed in \cite{NL1,NL2}. These lines deliver certain amount of signal to the neighboring antenna that counteracts for the coupling between them. The major drawback is huge amount of optimization involved in the adjustment of the width, length and the location of connection point of these lines. More than one neutralization line is needed to improve the isolation over wide frequency bandwidth and this complicates the process further. Parasitic elements also uses the idea of field cancellation to enhance the isolation between the antennas \cite{PE1,PE2}. The major draw back is the optimization of the shape and the dimensions of parasitic element that play significant role in isolation enhancement.

In \cite{DGS1} a periodic S shaped DGS was used to improve the isolation between two patch antennas to $23 \enspace dB$. Two DGS (T-shaped and Line slot) were used to improve the isolation to $18 \enspace dB$ over a wide frequency range of 3.1-10.6 GHz in \cite{DGS2}. Two DGS were used to address two different frequency bands. A DGS etched in a square ring fashion was used to improve isolation by $7 \enspace dB$, for a square patch surrounded by a square ring patch design in \cite{DGS3}. An isolation was improved to $17 \enspace dB$ for a $2$ element, $4$ shaped dual band design, where the length of the rectangular slot and spirals in the DGS were used to tune the frequency band in \cite{DGS4}. In \cite{DGS5} an isolation of $15 \enspace dB$ was achieved for a $2$ element MIMO antenna by combining two isolation enhancement mechanisms that are orthogonal placement of antennas and introducing a slit in the ground. Isolation enhancement for a MIMO antenna system for more than $2$ elements is difficult. Isolation was enhanced to $12 dB$ for a very closely packed $4$-element MIMO design, where they used four wideband DGS with multi objective fractional factorial design \cite{DGS6}. A slitted pattern was etched between $2$-element PIFA and wire monopole design that enhanced the isolation to reach $20 \enspace dB$ \cite{DGS7}. Moreover the study was extended to $4$-element PIFA (aligned along a line) to get an isolation of $12 \enspace dB$. Two DGS were used in \cite{DGS8} to improve isolation of a $4$-element MIMO design. A rectangular slot and a stair case slot were used to enhance the isolation to $12 \enspace dB$ between the horizontal and verticals antennas respectively. For an $8$-element MIMO design a very complicated DGS that consist of closed loop frequency selective surfaces and quad strips connected with a circular arc were used to improve isolation to $15 \enspace dB$ respectively \cite{DGS11}. 

The main problem in the DGS method to enhance isolation is the shape, size, number of DGS, position and the huge amount of optimization involved in the placement of DGS. We observed that from the perspective of EMP there is no systematic method to enhance the isolation, all of the available literature rely on the past experience and different parametric studies to obtain the enhanced isolation. This was the reason that the TCM was used for the isolation enhancement to develop a proper methodology for isolation enhancement.

\section{Isolation Enhancement using TCM}
The presence of the any antenna or deformation in the chassis greatly affect the chassis modes \cite{Asim_MOTL,Asim_EuCAP_2017_1,Asim_APS_2018_1,Asim_APS_2018_2,Asim_APS_2017_2}. TCM was used in  \cite{TCM_25,TCM_28,TCM_32,TCM_30,TCM_31} to enhance the isolation between the MIMO antenna designs. It was demonstrated in \cite{TCM_25} that the current in PIFA is more localized as compared Monopole and this is the reason that PIFA antennas are narrow band as compared to Monopole antennas. At the same time MIMO PIFA antenna design has better isolation because of its localized current nature. For frequencies less than 1 GHz, the chassis starts contributing to the antenna performance because now the electrical length of the antenna is huge. So for a MIMO antennas the situation will worsen because now all the antenna will excite the chassis modes. For frequency of 1 GHz, the antenna is having only one chassis mode having electric field maxima at the edge and electric field minima at the center. An electrical antenna placed at the edge will effectively excite the chassis modes while antenna placed at the center will not. Antennas placed at the mentioned position achieved a 5 dB more isolation as compared to the antennas placed at the edges of the chassis \cite{TCM_25}. A magnetic and electrical antenna placed at the edges will have improved isolation because the electrical antenna will excite chassis currents while the magnetic antenna will not \cite{TCM_28}. Co-located antennas were introduced in \cite{TCM_28}, to have very compact MIMO antenna design. To have better isolation, the co-located Magnetic and electric antenna excite the chassis currents in opposite direction. In \cite{TCM_25,TCM_28}, only the behavior of the first mode using TCM was observed and analysis were based on it.

The selective excitation of characteristic modes that have orthogonal behavior can enhance the isolation. In \cite{TCM_69}, a feed network (consisting of 4 hybrid 180 couplers) and 4 capacitive coupler were designed to excite the four different modes of the antenna. The modes excited due to its orthogonal nature results in an enhanced isolation. The asymmetry in the ground plane produces a natural tilt and increase in directivity \cite{TCM_70}. Two antennas were placed at asymmetrical ground plane such that both of them excite different characteristic modes, it will help in getting highly isolated and highly uncorrelated beams. 

In \cite{TCM_72}, the lower order modes were separated by the use of decoupling network and GA was used to synthesize low Quality factor MIMO antenna. Quality factor is inversely proportional to the antenna bandwidth. A monopole (excited via CCE) and chassis (excited via ICE) combination was used to implement highly isolated MIMO design \cite{TCM_58}. From the combination of CM, a new set of Radiation modes can be formed. The radiation modes formed are highly orthogonal to one another thus resulting in highly uncorrelated beams \cite{TCM_74}.

In \cite{TCM_73}, out-of-band interference was improved by the use of TCM. In Aeroplane for communication they use antenna at 2.8 to 24 MHz with very strong power, the harmonics of this antenna are also high power and thus affects the communication of other antennas. Such type of interference is called out-of-band interference. The Methodology adopted was to  calculate the Modal mutual admittance (MMA) and Modal Self admittance of the antenna. CMA analysis of the higher frequency antenna (with low power) is calculated. Functional and non-functional modes of the first antenna is calculated. Modes contributing to real communication are known as functional while the modes contributing to interference are known as non-functional. Non-functional modes are blocked by inductor loading with the help of TCM. The value of the inductor shall be properly optimized to block the non-functional mode.

In \cite{TCM_71}, designs made with the application of TCM were compared with the empirically made designs. 5 designs were opted from different papers with 2 designs opted from TCM and 3 opted by empirical method. All of the selected designs were MIMO design and they were compared in 7 different real time scenarios. The scenarios were that first all the designs performance was observed in the presence of a box then 3 scenarios in which the antenna was hold by one hand and then 3 scenarios in which the antenna was hold by two hands. TCM design recorded 3 times high multiplexing efficiency as compared to conventional design and for all the 7 scenarios, TCM designs performed with ME of 1.6 dB better.

Now all these methods focus on the isolation enhancement for frequentness around 1 GHz \cite{TCM_25,TCM_28}. At 1 GHz, for a normal chassis only one chassis mode is present. The problem will escalate, if we consider frequency greater than 1.5 GHz as now more than one mode will be present. As the point where one mode was having current minima (a possible location for the placement of the second antenna), another mode current maxima lies over there. Thus we are left with no possible location.

\begin{figure}[h]
	\begin{center}
		\noindent
		\includegraphics[width= 8.75cm]{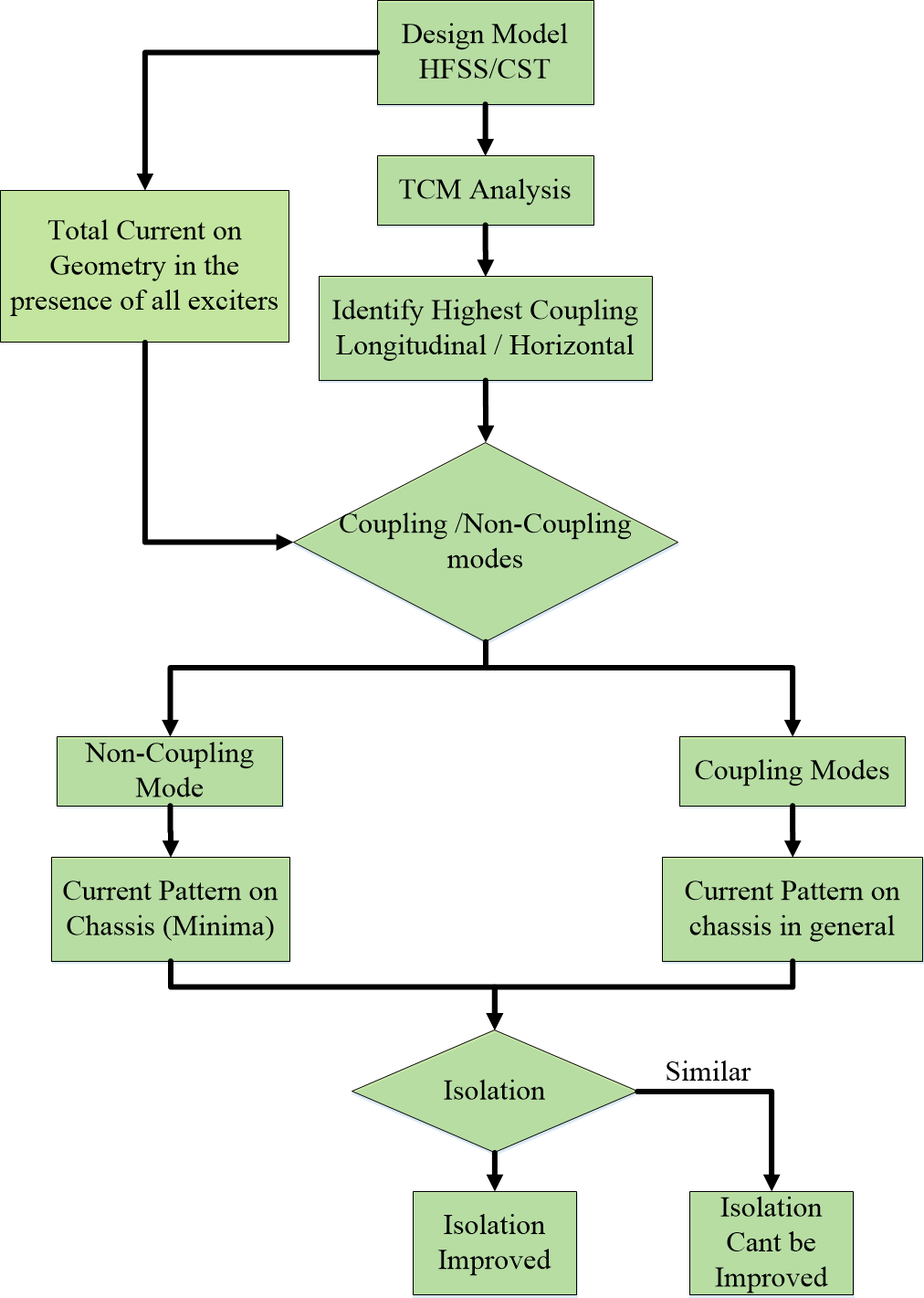}
		\caption{Block diagram of Proposed Isolation enhancement scheme that can predict whether the isolation can be enhanced or not \cite{Asim_Access}}
		\label{Block_diagram}
	\end{center}
\end{figure}

In \cite{Asim_Access}, a possible solution to this problem was proposed, where the DGS was used to enhance the isolation between two MIMO antennas. A methodology was also proposed to predict whether the isolation can be enhanced or not. The block diagram of the methodology is shown in the {Figure} \ref{Block_diagram}. After the designing of MIMO antennas the characteristic modes were identified. The modes were classified as coupling and non-coupling modes based on the current distribution and its resemblance with the chassis current distribution in the presence of the excitation sources. Non-coupling modes are the one contributing to the radiation while the coupling modes are contributing to the port coupling thus degrading the isolation performance. If there exists a certain location on the chassis, where the DGS placed can block the coupling mode but at the same time does not effect the non-coupling mode the isolation between the antenna elements can be enhanced but if there does not exist any such location, the isolation between the antenna elements cannot be enhanced. The method was applied to different antenna designs and it achieved the most significant enhancement in the isolation as compared to other designs in the literature.

\section{Conclusion}
In this paper, a summary of different available techniques to enhance the isolation between MIMO antenna elements is proposed. The shortcomings of the different EMP based approaches is discussed. The TCM based approaches to enhance the port isolation between the MIMO antenna elements are also discussed. It was lastly shown that a new method to enhance the isolation between the MIMO antennas with the help of TCM was proposed. The method is able to predict whether the isolation can be enhanced or not.

%
%

%
%
%
%
\end{document}